Australia’s National
Science Agency

# Applying causal inference to inform early-childhood policy from administrative data

Elena Tartaglia[1] and Peter Rankin[2]

10 January 2023

1. Commonwealth Scientific and Industrial Research Organisation, Australia.

2. Queensland Brain Institute, University of Queensland, Australia.

# Contents

**Abstract**

Improving public policy is one of the key roles of governments, and they can do this in an evidence-based way using administrative data. Causal inference for observational data improves on current practice of using descriptive or predictive analyses to inform policy decisions. Causal inference allows analysts to estimate the impact a policy change would have on the population if the encoded assumptions about the data generation process are valid. In this paper, we discuss the importance of causal analysis methods when analysing data to inform policy decisions. We take the education sector as a case study and provide examples of when to use a causal analysis. We use simulation to demonstrate the vital role causal diagrams play in variable selection and how bias can be introduced if extraneous variables are included in the model. Our exploration provides clear evidence for the utility of causal methods and practical examples of how to conduct such analyses. The paper promotes the incorporation of these methods in policy both for improved educational outcomes and scientific understanding.

Keywords: causal inference, administrative data, DAGs, policy development, early childhood development

**Acknowledgements**

The authors would like to thank Christopher M. Baker, Robert Dunne, Simon Knapp, Marc Wigzell and the First Five Years team at the Department of Education, Skills and Employment for useful discussions. This research was supported (partially or fully) by the Australian Government through the Australian Research Council's Centre of Excellence for Children and Families over the Life Course (Project ID CE200100025) and the Data Integration Program for Australia.

# 1 Introduction

Improving educational outcomes through evidence-based policy is an important but complex task undertaken by governments across the globe. Successful investment in education has large returns for human capital development and well-being of citizens. Yet, the educational landscape is complex with multifaceted outcomes and measures of input that are the product of interactions between individuals, communities, culture, and legislation. Policy makers thus rely in part on evaluations of new interventions to guide decisions.

Analysis of observational, administrative data is frequently used to assist policy makers by providing evidence-based evaluations. Governments collect extensive administrative data across various departments, and these datasets may be linked to evaluate the effects of policy changes. The ease of these analyses has markedly increased with open-source and user-friendly packages in programming languages such as R and Python. Likewise, hardware is more affordable and the number of analysts has increased (Almgerbi et al. 2022; Henry and Venkatraman 2015), making data analysis a viable option to assist policy makers.

Despite the potential of observational data analysis to provide useful evaluation, confounding, non-random sample selection, and unobserved factors frequently constrain the ability to infer causality of the impact of policy changes on the population. Making this causal inference is crucial as it provides the link between the policy change and outcome. The typical solution around confounding and selection is to design and implement randomised control trials. Briefly, by randomising who undergoes the policy change (intervention condition) and who does not (control condition) the resulting comparison of outcomes is less likely to be biased by confounding and selection. This greatly strengthens the inference of causality. However, administrative data is generally non-randomised meaning that the intervention and control conditions generally differ in meaningful ways other than exercising the intervention, so comparing them directly may not indicate a difference caused by the policy change.

Causal inference techniques for observational data have been developed to specify assumptions about how the data was generated and then use statistical techniques to remove the biases that these introduce (Pearl 2009; Hernán and Robins 2020). If the assumptions are met, causal inference can be claimed with greater certainty. One way to represent data generation process of observational data is via causal diagrams using directed acyclic graphs (DAGs) (Pearl 1995). These graphs encode the assumptions, typically informed by prior knowledge, about how the data was generated. Based on these data generating assumptions, the causal diagram can then be used to select the adjustment variables which remove bias from a statistical estimate of the causal effect.

Though this process occurs informally in most applications of statistical adjustment (e.g., linear regression), a common misconception regarding this process is that all variables related to the intervention or outcome need to, and should, be included. However, causal diagrams can be used to identify the variables that should be adjusted for and indicate the variables that should not be included, even though they are linked to the treatment and/or outcome.

In this paper, we aim to demonstrate how the naïve application of statistical techniques for observational data can produce misleading results if not properly situated within a causal diagram. Further, we show how advances in causal inference methods beyond simple linear regression can be applied to better avoid these pitfalls in an education context. Our demonstrated use of DAGs and relatively new statistical adjustment methods (Hernán and Robins 2020) can help analysts strengthen the evidence base available to policy makers.

The paper first introduces causal inference and the theory of causal diagrams. It then presents a simulated case study of the impact of childcare attendance on conduct issues of children in school. The simulated data demonstrates how, in contrast to common practice of including extraneous covariates, adjusting for the necessary variables determined by the causal diagram better estimates the causal effect. We conclude by discussing practical challenges when applying these methods.

# 2 Background on drawing causal inference

Causal inference techniques for observational data aim to reveal if, and the extent to which, a change in one variable causes a change in another. For example, whether attending childcare causes an increase or decrease in the language and cognitive development of children. Ideally, such questions would be answered through a Randomised Control Trial, where subjects are assigned to the treated group (attending childcare) or control group (not attending childcare) at random. Though researchers, governments, and educational stakeholders are interested in answers to such questions, they are often limited by resource constraints or the practical and ethical implications of randomisation. Governments and researchers may, however, have access to detailed administrative data on a topic of interest. This data is known as *observational data*, as it may record observations, for example, of whether children are participating in childcare to facilitate a comparison with those who did not.

Observational data is generally inappropriate for drawing causal inference as it contains underlying biases in terms of sample selection and unmeasured confounding. Causal inference techniques provide a method for reducing these underlying biases. Crucially, by generating causal diagrams, assumptions can be specified about how the data was generated and then, based on those assumptions, appropriate statistical adjustment methods can be used to adjust for variables identified as confounding the causal association. These “goldilocks variables” determined by the causal diagram are only the variables which remove bias and, crucially and at odds with common practice, excludes variables that inadvertently increase bias.

The causal diagram is made using prior knowledge in the form of existing literature, observations, and consultation with subject matter experts. This is necessary to provide insight into the causal relationships between the variables in the analysis and for specifying the assumptions of how the data was generated. Ideally, the processes generating the data would be known and the extent the observational data captures this could be evaluated, and additional information collected to remove remaining bias. However, educational systems and developmental outcomes are complex, and it is often impractical to know for certain how the data was generated. Thus, in many cases the casual diagram will represent a best estimate of the data generating process based on domain expertise. The process is immensely valuable, however, as proposed causal inferences can be evaluated in context of the diagram’s assumptions. As insight accumulates new diagrams can be proposed that build on existing bodies of knowledge and further improve the likelihood of making correct causal inferences.

In this section, we discuss the various types of research questions that come about in data analysis and discuss when causal inference techniques are appropriate. We give a basic overview of the theory of causal diagrams. Finally, we talk about the types of bias that occur in observational studies and how to identify them using causal diagrams.

## 2.1 When to use causal inference techniques

Before undertaking a statistical analysis, it is important to determine what type of research question is being asked, as the statistical techniques applicable vary depending on the question type. In particular, the method for selecting variables differs depending on whether the research question is *descriptive, predictive* or *causal* (Hernán, Hsu, and Healy 2019). This paper focuses on the techniques to be used in the latter case, but here we describe all three types of questions to allow data analysts to understand when causal techniques are appropriate for their study.

Whether a question is descriptive, predictive or causal relates more to how the results of the study will be used, rather than the question itself. Descriptive questions seek to determine whether there is an association between two variables in the population of interest. A predictive question seeks to predict one variable from another and are often useful for identifying or screening people at risk of a certain outcome. Causal questions seek to determine how changing one variable will impact another variable. The research question "does maternal mental health associate with child social and emotional development?" (Connell and Goodman 2002) could be thought of as either of these types of questions, depending on what application the research results will have. It could be thought of as a descriptive question in a preliminary analysis, leading to further questions about what causes the association. It could be thought of as a predictive question if the aim were to use maternal mental health to identify children in need of extra support. It could also be thought of as a causal question if the aim is to determine whether improving maternal mental health would improve the child's development.

When determining whether a research question is causal, it is important to think about whether the intervention is meaningful. Let us take the example of a research question to determine whether ethnicity is associated with high school completion. This could be thought of as a descriptive question in an initial exploration of the dataset or as a predictive question if the aim were to identify variation in high school completion rates for the allocation of extra support. This question, however, is not thought of as a causal question since it does not make sense to think of ethnicity as something that could be changed for a person. In other words, the intervention "ethnicity" is not meaningful.

The previous example where the "intervention" is maternal mental health, falls into the grey area in which many public policy questions fall. It is not clear how one would do the Randomised Control Trial to assign some mothers with mental health issues and some without. However, if the aim of the study is to understand whether improving maternal mental health would improve their child's development, this question is causal and causal inference techniques should be applied to adjust for selection and confounding biases in the observational dataset.

## 2.2 Individual and average causal effect

We have so far applied the word causal loosely. To improve precision, we now introduce the concepts of individual and average causal effects. Ideally, we would like to measure the individual casual effect: if I send my child to childcare, will that improve their school readiness? However, the individual causal effect is impossible to measure, as for each individual, we either have to do the intervention or not (either send the child to childcare or not). Instead of measuring the individual causal effect, we can measure what is known as the average causal effect in a population. The average causal effect describes whether there was an average change in the outcome due to the intervention, at a population level. It answers questions like "does sending children to childcare on average improve their school readiness?"

There are two ways to measure the average causal effect, known as the *causal risk difference* and *causal risk ratio*. For a binary outcome, such as having good or bad conduct at school, the causal risk difference is the difference between the probability of having good conduct if children are sent to childcare with the probability of having good conduct if children are not sent to childcare. The causal risk ratio is instead the ratio of these quantities. A causal risk difference of zero or a causal risk ratio of one both indicate no causal effect. Both measures are valid, but for our examples we use the causal risk ratio, as it arises naturally from the models we are using.

## 2.3 Causal diagrams

A causal diagram is a way to specify assumptions about how the observational data was generated. Once these assumptions have been established, the causal diagram is used to determine which variables should be adjusted for to remove underlying bias in the data. To understand from the diagram which variables these are, it is necessary to understand some of the underlying theory of causal diagrams, so we give a brief introduction in this section. The causal diagrams in this paper were made using the software DAGitty (Textor et al. 2016).

A causal diagram is a directed acyclic graph (DAG) with nodes representing measurements and arrows representing average causal effects. It encodes assumptions made by an investigator based on their domain knowledge about how the observational data was generated. Figure 1 shows two simple causal diagrams: the leftmost depicts the assumption that the variable A has a causal effect on the variable B, indicated by the arrow pointing from A to B, and the rightmost depicts no causal effect, since no arrows connect the variables. A causal effect between A and B also implies the variables are associated. The direction of the arrow is not important when thinking about association, as it is equally true to say A is associated with B as B is associated with A. In this way, we can think of association as flowing through the arrows of a causal diagram, regardless of the arrow direction. However, we can think of an association between two variables as being causal if the path between those variables is traversed in the arrow directions. If when traversing a path between two nodes, any of the arrows are traversed in the wrong direction, then this association is noncausal. To remove biases between the treatment and outcome, we must *block* any noncausal association between the treatment and outcome. Choosing which variables to adjust for in an analysis to remove bias is equivalent to blocking any paths in the causal diagram between treatment and outcome through which noncausal association flows. A path through which association flows is called an *open* path and one through which it does not is *closed.*

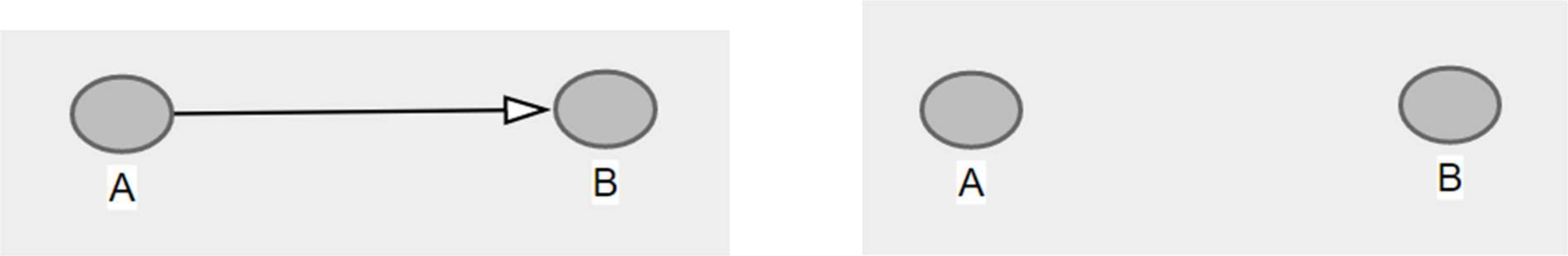

**Figure 1 The left causal diagram represents the assumption that variable A has a causal effect on the variable B. In the right causal diagram, the two variables are not associated, since there is no path connecting them.**

The structure of a causal diagram can be determined by breaking it into its three basic building blocks shown in Figure 2.

The first diagram shows variable C as a common cause of the variables A and B. Though there is no causal association between A and B (no path between A and B following arrows in the direction they point), the common cause C induces a noncausal association between A and B through what is termed a *back-door path* from A to B through C. This means that there is an open path between A and B, through C, which is not causal, because the arrow from A to C is traversed in the wrong direction. This noncausal association is a bias which can be removed by adjusting for variable C in the analysis. We say that the noncausal association between A and B is blocked by adjusting for the variable C, and that the path between A and B through C then becomes closed. For example, finances (C) of a childcare centre could be a common cause of the structural quality (A) of the childcare centre and the development of children (B) attending the centre. Specifically, childcare centres with a larger operating budget would have capital funding for improving physical aspects of the centre, but they could also retain high quality teachers through financial incentives or better working conditions. Consequently child development improves via the sustained high-quality interactions with the teachers (Rankin et al. 2022). For data generated in this way, there would be an association between the structural quality of the childcare centre (A) and the developmental skills of the children who attend it (B) which is not causal: it would be induced by the open backdoor path through the finances of the centre (C). However, adjusting for finances by only comparing centres in a similar financial situation would remove the association by closing the path through the variable finances.

The middle diagram shows a collider C, which is a common effect of the variables A and B. The collider C does not introduce any bias into the analysis, provided it is not included in the adjustment process. However, adjusting for the collider C, unblocks the path between A and B, inducing a noncausal association between the variables. In other words, the collider C blocks any noncausal association between A and B, meaning the path between A and B is closed, but adjusting for C would open the path. For example, quality of childcare C could be a common effect of child temperament A and wealth of family B. This situation could occur if children with a difficult temperament and those from wealthier families are more likely to attend high quality childcare (Davis, Eivers, and Thorpe 2012). Though child temperament and wealth of family are not associated in data that has a range of qualities of childcare (the path between them is blocked by a collider), if the data is restricted to only high-quality childcare, this would induce a negative association between child temperament and wealth of family (opening the path between them). This is because, looking only at high quality childcare, if a child has an easy temperament, they are likely to come from a wealthy family, and if they come from a low-income family, they are likely to be there because they have a difficult temperament. This induces the noncausal association that children from wealthier families have easier temperaments. Therefore, it is important to avoid adjusting for colliders as it introduces bias into a study, and it is important to think about when data implicitly restricts to a certain part of the population as this amounts to implicitly adjusting for a variable.

Finally, the rightmost shows a mediator C between variables A and B, representing the situation where A has an average causal effect on C which has an average causal effect on B. An example of this could be teacher training A affecting teacher-student interaction C which in turn has affects their student's cognitive development B (Egert, Fukkink, and Eckhardt 2018). Provided the mediator C is not adjusted for, there is a causal association between A and B as the path between A and B is open. Instead, adjusting for the variable C in the analysis would block the association between A and B, so the path would be closed. For our example, if the data was generated according to the rightmost diagram in Figure 2, then there would be an association between teacher training and their student's cognitive development. However, if the data were adjusted for the mediator teacher-student interaction, by looking within levels of teacher-student interaction, there would no longer be an association between teacher training and their student's cognitive development. By adjusting for the mediator teacher-student interaction and closing the path, the association between teacher training and their student's cognitive development would no longer be present.

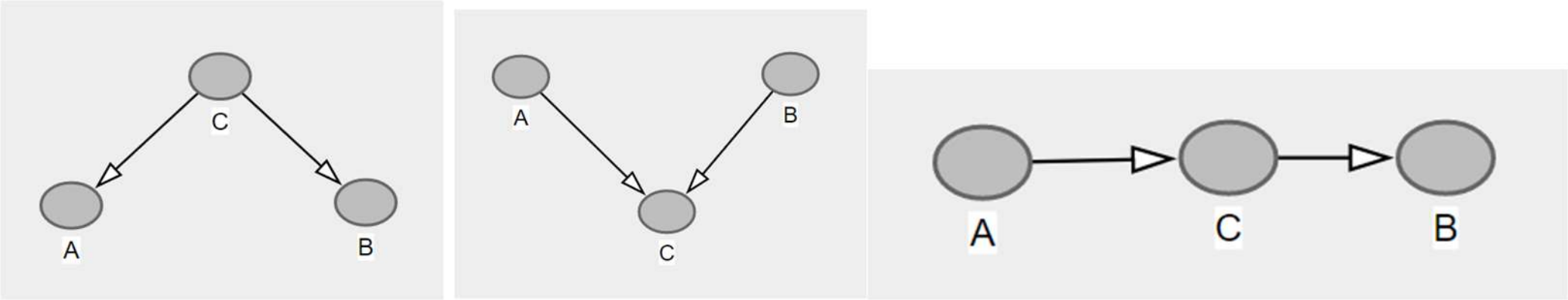

**Figure 2 Basic causal diagrams. In the leftmost causal diagram, the variable C is a common cause of variables A and B. This means there is an association between A and B that is not causal, through the back-door path through C. In the middle diagram, the variable C is a collider, a common effect of the variables A and B. There is no association between A and B in this diagram but adjusting for C would induce a noncausal association between them. The rightmost diagram shows a mediator C between the variables A and B**

A full causal diagram is built starting from the treatment and outcome, adding nodes for their common causes and then repeatedly adding common causes of any nodes in the graph. Common effects should be added if they are implicitly adjusted for in the analysis: for example, if all individuals in the data set have the same value for one of the variables. Mediators are generally only included to help block noncausal associations caused by unmeasured variables which can therefore not be adjusted for directly to remove bias.

In the next section, we will use these diagrams to describe the types of bias that can occur in observational studies. The building block diagrams of Figure 2 and the understanding of how to open and close paths to the flow of association will be key tools in understanding how to adjust for the types of bias. See Appendix A.1 for simulated examples demonstrating open and closed paths in the three building block diagrams.

## 2.4 Types of bias and choosing variables to adjust for them

There are three main types of bias that need to be addressed in observational studies: confounding, selection and measurement bias. Causal diagrams are used to identify the bias and select which variables should be adjusted for to remove bias. The causal diagrams in this paper have been made using the software DAGitty (Textor et al. 2016), with its default colour scheme (e.g. green with a triangle for treatment, blue with a bar for outcome and white for controlled variables).

Confounding bias occurs when there is one or more common causes of the treatment and outcome. These common causes create a back-door path: an open path between the treatment and outcome through which noncausal association flows. Since this association does not describe a causal effect, it introduces bias when estimating the effect of the treatment on the outcome.

An example of confounding bias is shown in Figure 3. The research question seeks to understand the impact of structural quality of the learning environment on the cognitive, social and emotional development of a child. If the data was generated according to the relationships in Figure 3, there would be no causal effect of structural quality on the development on the child. However, the data would show an association between the two, since centres with more money can afford greater structural quality and attract more educated and experienced teaching staff who improve process quality at the childcare which improves the development of the child. Looking at the pure association in the data would show a positive association between structural quality and child development, but this relationship is not causal. The association comes from the open back-door path between structural quality and child development through finances. Closing this path by adjusting for finances, for example by comparing centres that have similar funding, would remove this bias and show no association between the structural quality and the cognitive, social and emotional development of the child. In general, confounding bias is removed by adjusting for variables to close back-door paths between the treatment and outcome.

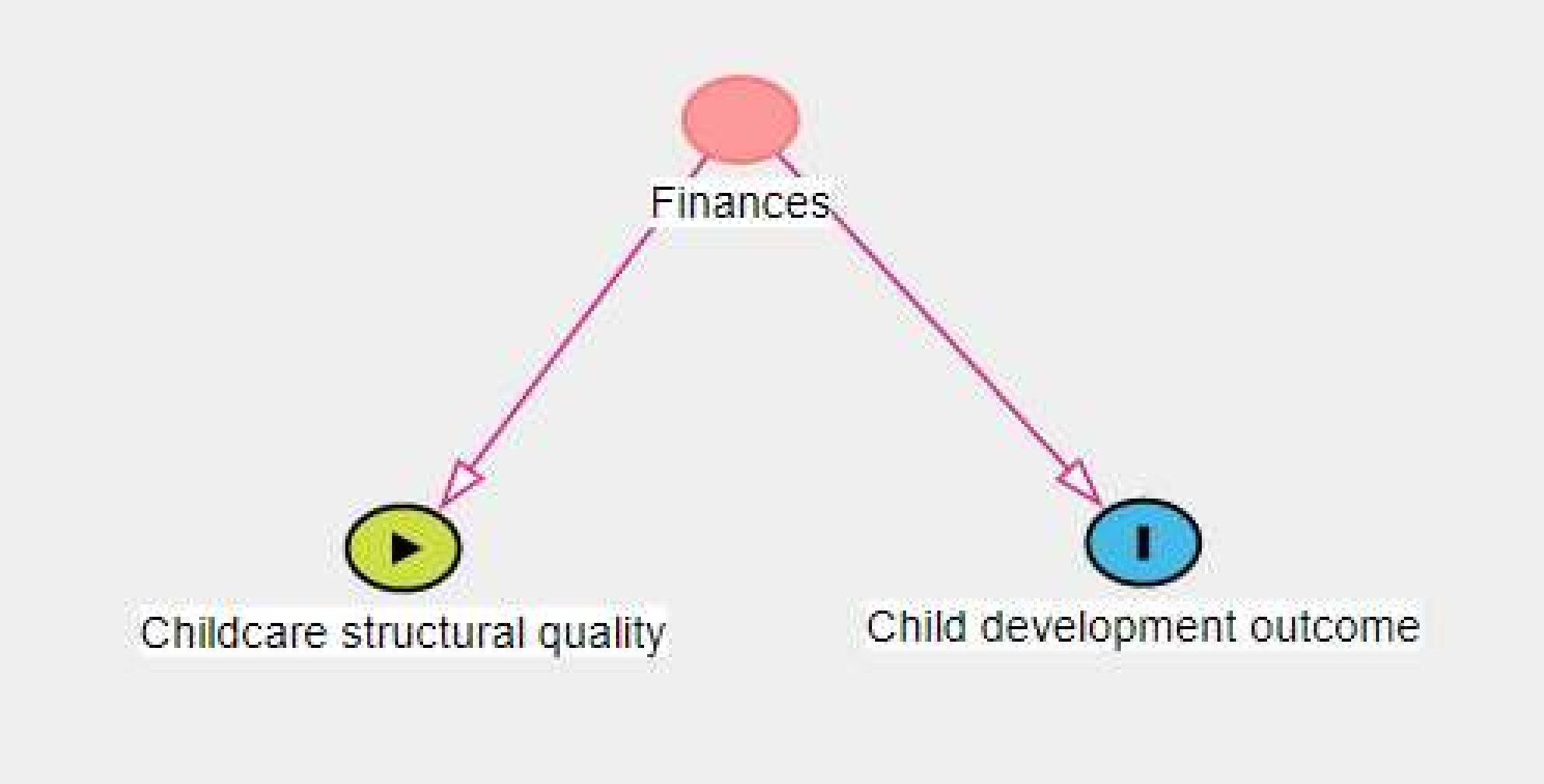

**Figure 3 Confounding bias. When looking at the impact of structural quality of the learning environment and cognitive, social and emotional development, a common cause could be the finances available to the childcare facility. If centres with more money to spend have greater structural quality but also provide higher process quality childcare by being able to employ more educated and experienced staff, the data would show an association between childcare structural quality and cognitive, social, and emotional development which is not causal. In the situation represented by this diagram, any association in the data between structural quality of childcare and developmental outcomes would be removed by adjusting for finances, since there is no causal effect.**

Selection bias creeps into an analysis by conditioning on a common effect (collider) of the treatment and outcome, or a direct effect of such a collider. The name selection bias comes about because it often occurs when a collider variable is adjusted for implicitly by selecting for a certain part of the population. Since adjusting for a collider opens a path to the flow of association, this introduces non-causal association between the treatment and outcome, leading to a biased estimate of the average treatment effect on the outcome.

For example, studying the impact of hostile parenting (negative affect or indifference to child, including use of coercion, threat, or physical punishment; (Lovejoy et al. 1999)) on an externalizing behaviour (aggression, conduct problems, hyperactivity; (Pinquart 2017)) in their child, a common effect could be whether the children and parents had attended a parenting intervention (such as a class to learn new parenting techniques), Figure **4**. This could be because parents with hostile behaviours may be more likely to want to improve their parenting and attend the intervention. Further, parents with children who have behavioural problems may want to attend the intervention to improve their parenting. The collider variable “parenting intervention,” does not cause bias between the treatment and outcome, since colliders form closed paths, blocking the flow of association. However, if for example, the data were only available for children who attended the intervention, perhaps because the data came from a convenience sample of parents attending the intervention, then the data implicitly adjusts for attending the intervention. This introduces bias by opening the path from treatment to outcome through the collider, inducing a negative association which is not causal. We can see this as follows: by restricting to children who attend the intervention, the children in the dataset are likely to either experience harsher parenting or have greater externalizing behaviour problems. This means that children who experience harsher parenting are less likely to also have externalizing behaviour problems, which leads to a negative association which is not causal: the data would incorrectly show that children who experience harsh parenting have better behaviour. Therefore, it is important to think about which cohorts have been excluded from the data analysis and determine whether this causes selection bias.

More generally, selection bias refers to all biases from conditioning on a common effect of two variables, where one is either the treatment or a cause of treatment, and the other is either the outcome or a cause of the outcome. Figure 5 gives an example of bias introduced by adjusting for a collider between the treatment and a cause of the outcome. In general, adjusting for a collider on a back-door path between treatment and outcome will open the path and introduce bias into the estimate of the average treatment effect.

Measurement bias arises when the data does not contain a variable we need to adjust for, but it includes something similar. An example could be that while socioeconomic status is a common cause of attending childcare and language and cognitive development, the dataset only contains an estimate of a child's socioeconomic status based on where they live. This can be incorporated into the causal diagram by adding separate nodes for the unmeasured true variable and its measured proxy, drawing the appropriate arrows to other nodes. Further details on measurement bias can be found in (Hernán and Robins 2020).

Causal diagrams are useful for deciding which variables to adjust for, since failing to adjust for common causes will lead to confounding bias, whereas adjusting for colliders or mediators can introduce bias. Blindly including all variables in the observational dataset can lead to biased results, so it is crucial to use a causal diagram to record assumptions about the data generation and from that determine which variables should be adjusted for. It is important to understand what populations have been implicitly selected for in the data collection process to determine whether selection bias is occurring. Adjusting for a mediator between the treatment and the outcome can also lead to bias, as it will block some of the causal association, as shown in Figure 6.

In this section we have shown how to use causal diagrams to identify which variables need to be adjusted for to remove bias. In the next section, we give a simulated case study, detailing how to create the causal diagram and statistically adjust the data to obtain an unbiased estimate of treatment effect on the outcome.

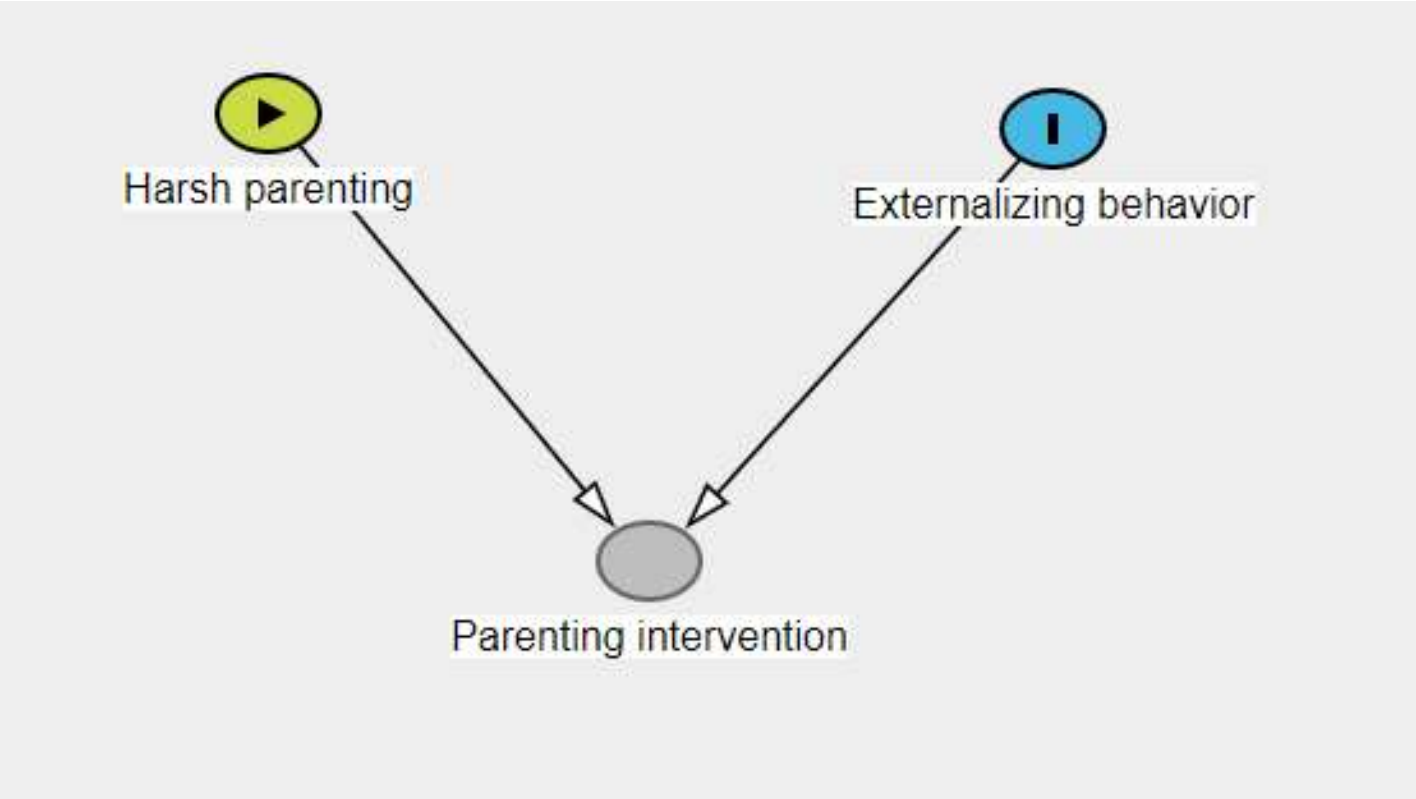

**Figure 4 Selection bias. Imagine an observational study that uses a cross-sectional convenience sample to evaluate whether harsh parenting is associated with externalizing behaviour. This sample is drawn from a parenting intervention. In the scenario, a) parents who use harsh parenting are more likely to be enrolled in the intervention, since they are likely to want to learn new parenting techniques, and b) parents whose children have high levels of externalizing behaviour are also more likely to be enrolled in the intervention, because their children behave contrary to their parents' expectations and parents would like to improve their behaviour. Since the data is restricted to those who participate in the intervention, there could be an observed negative correlation between harsh parenting and externalizing behaviour which is not causal (harsh parenting would seem to improve a child's behaviour). This is because children who enrol in the intervention are likely to have experienced harsh parenting or have higher externalizing behaviour problems, so if they experience harsh parenting, they are unlikely to also have externalizing behaviour. A situation like this leads to selection bias in the average causal effect estimate.**

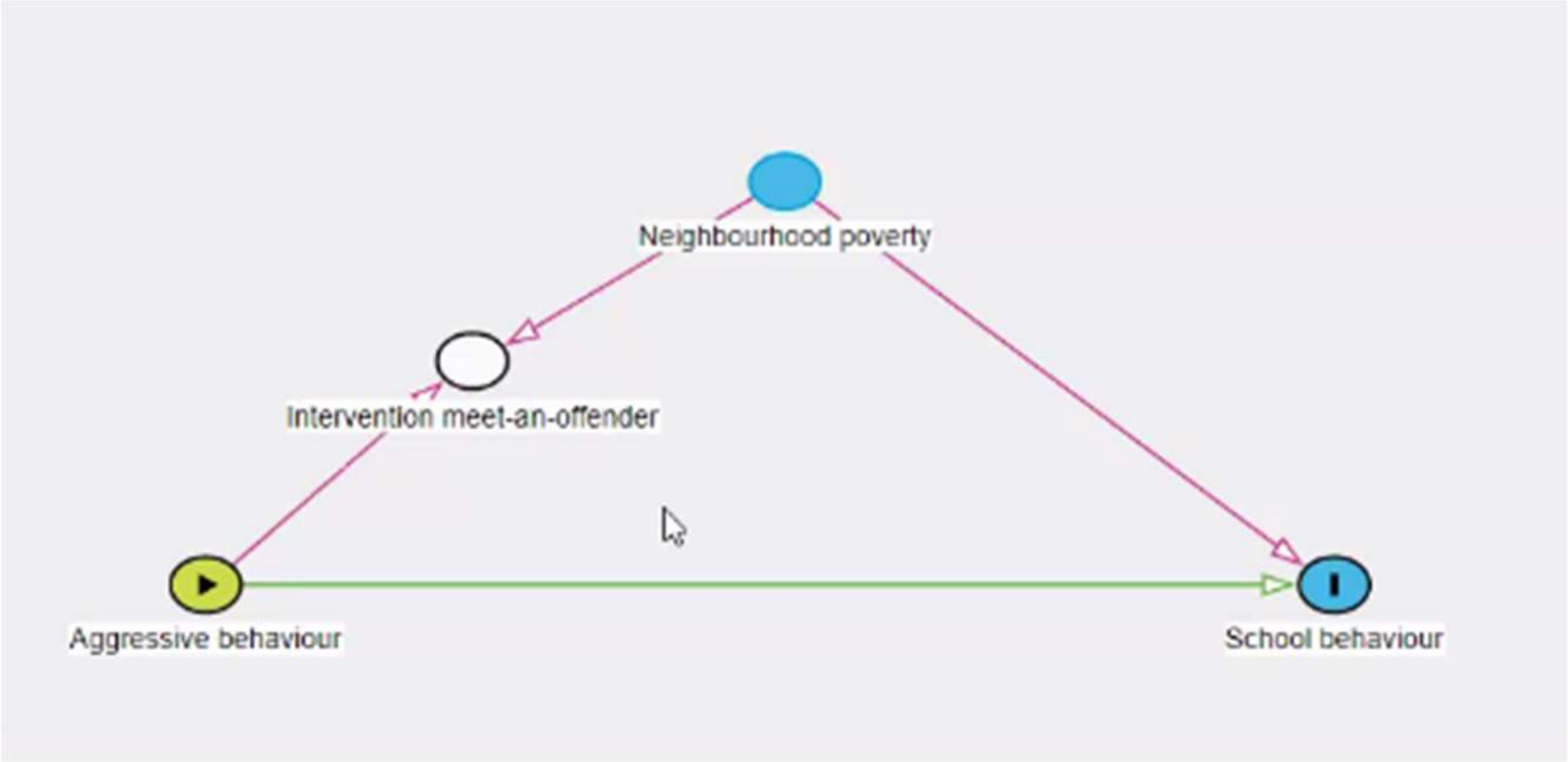

**Figure 5. Selection bias. Adjusting for a collider between the treatment A and a cause of the outcome Y can also introduce selection bias. This example seeks to look at the impact of aggressive child behaviour on their behaviour in school, where the data has been taken from a group of children who attended a meet-an-offender session. Assuming that the two main reasons for children to attend the session are that they have exhibited aggressive behaviour or they live in a poor neighbourhood, by only looking at children who attended the meet-an-offender session, the data would open the path between aggressive behaviour and neighbourhood poverty, through having attended a meet-an-offender session. This noncausal association can be removed, however, by also adjusting for neighbourhood poverty to close the path.**

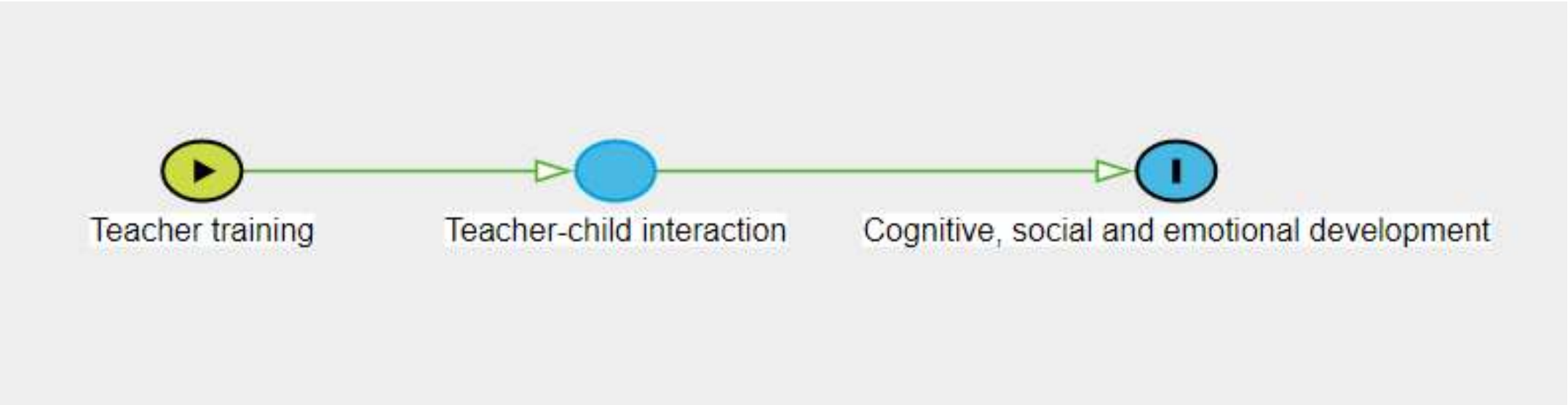

**Figure 6 Mediator. In this example, teachers participating in a training intervention (e.g., Egert et al., 2018) improves children's cognitive, social, and emotional development only through the improvement in interaction quality with the children. Adjusting for interaction quality, by including it in the model, would remove any association in the data between the teacher training intervention and cognitive, social, and emotional development, making it seem as though there were no causal effect of the intervention, even though there is one.**

# 3 Case Study

In this section we present simulated case study where we ask whether childcare attendance improves the behavioural development of children. The purpose of this simulated case study is two-fold. The first is to give a worked example of how to draw causal inference from observational data, from drawing the causal diagram to doing the statistical analysis and interpreting the results. The second is to give a concrete example of how bias can be removed and introduced by either including or excluding certain variables. Furthermore, we demonstrate how selection bias can occur implicitly when the data is restricted to a certain part of the population, as well as how this bias can sometimes be removed. We hope this practical guide not only demonstrates the value of drawing causal diagrams when analysing observational data but gives the reader the tools to try it on their own datasets. Code in R will be available in the supplementary materials.

In this section, we explain how to build causal diagram to answer the research question "does childcare attendance improve the behavioural development of children?" In this hypothetical study, we imagine we obtained the data from a weekend playgroup.

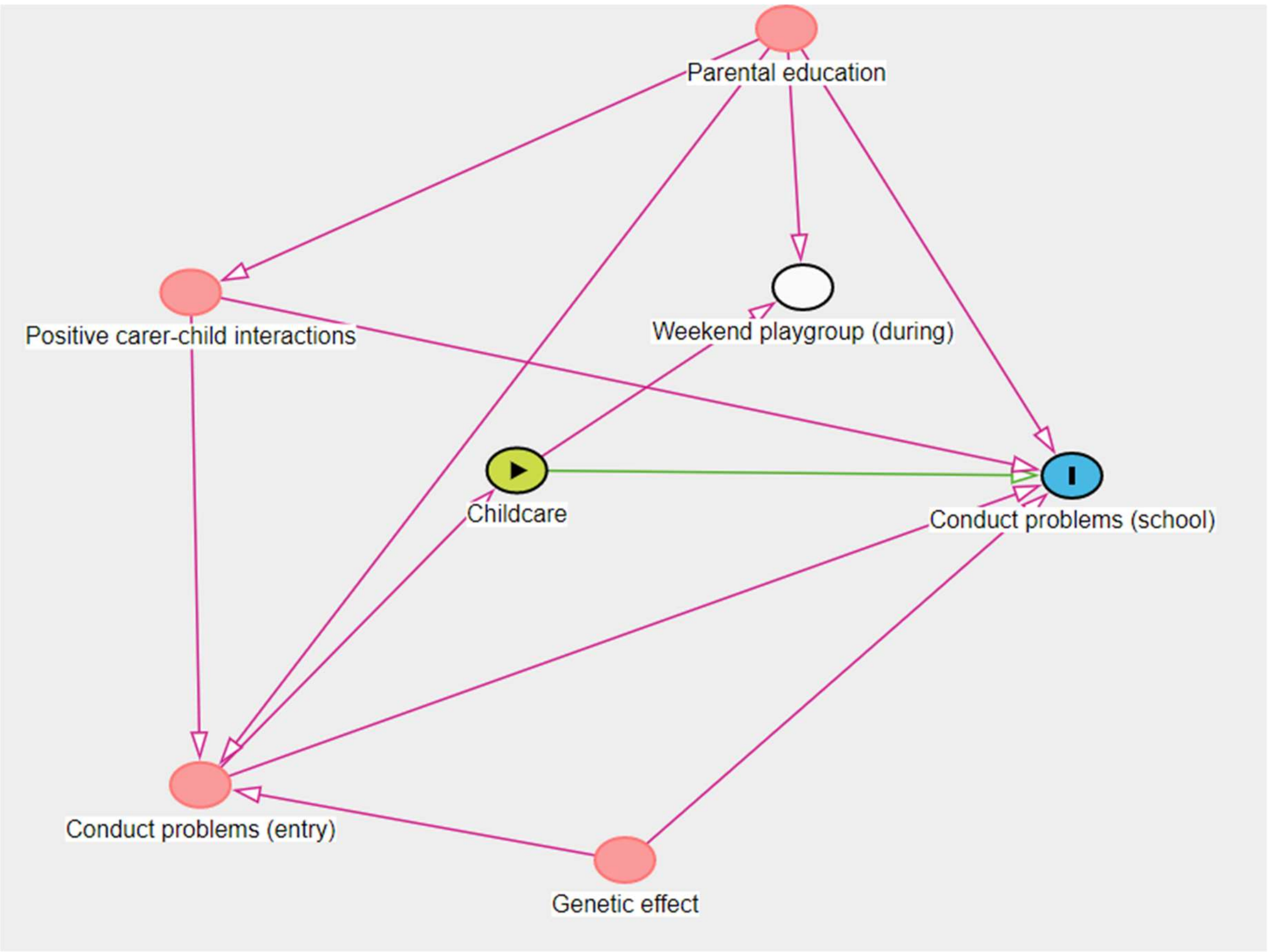

**Figure 7 Causal diagram for case study. The research question is "does attending childcare improve a child's behavioural development?" The intervention is childcare, and the outcome is conduct problems in school.**

To answer the research question "does attending childcare improve a child's behavioural development," we seek to estimate the causal effect of attending childcare on whether the child has conduct problems at school. Therefore, the treatment is a binary variable indicating whether a child attended childcare or not, and the outcome is a binary variable indicating whether the child has conduct problems at school or not. Though causal inference techniques apply to binary, categorical and continuous variables, for simplicity, we will only use binary variables in this case study.

To build the causal diagram, Figure 7, we follow the procedure suggested in the edX course (Hernán 2019)

1. Start with treatment and outcome, drawing an arrow from treatment to outcome.
2. Add common causes of treatment and outcome and repeat, adding the common causes of any nodes in the graph.
3. Add selection nodes: any nodes that are colliders that have been conditioned on.
4. Add measurement error nodes.

In step one, we add the treatment node "childcare" and outcome node "conduct problems (school)," drawing an arrow from "childcare" to "conduct problems (school)," because we're interested in measuring the effect of childcare on a child's behavioural development, which we're measuring based on whether they are recorded as having conduct problems in school.

In step two, we add common causes of these two nodes: conduct problems on entry, which describes whether the child had pre-existing conduct problems before starting childcare. This is because a child having behavioural issues can impact whether they attend childcare and also impact whether they have conduct problems at school. We then proceed to add nodes for any common causes of the current nodes: genetic effect, parental education and positive carer-child interaction.

Behavioural development at school entry is caused by the interplay between the child's genetic profile, pre-existing development, and the environment (Rutter 2006). Therefore, the child's existing behaviour, in the form of conduct problems at entry to childcare, is our first common cause. It is expected to impact a child's childcare attendance and their conduct problems at school (May et al. 2021). The conduct problems themselves are caused by the child's genetic profile (interacting with the environment), so we include it as their common cause (Weeland et al. 2015). Next, we turn our attention to important environmental effects, adding positive child-carer interactions as a common cause between the two conduct problems variables (Pinquart 2017). Parental education is also thought to have an impact on a child's behaviour (Letourneau et al. 2013) as well as their child-carer interactions (Moon, Damman, and Romero 2020), so we add it as a common cause of those three nodes.

Having included our important common causes, for the purposes of this simplified example, we move to step three where we add selection nodes. In our hypothetical example, we imagine having collected data from children who attend a weekend playgroup. In our example, the weekend playgroup is advertised at our childcare centre and in areas frequented by parents with higher formal education (such as libraries, theatres or museums), so both of these variables impact whether a child attends the weekend playgroup and, therefore, is part of our dataset. Since the variable "weekend playgroup" is a collider on the path between the treatment "childcare attendance" and outcome "conduct problems (school)", only having children who did attend the weekend playgroup in our dataset is an example of selection bias.

Finally, step four is to add measurement nodes, which are nodes to represent the variables we actually measured, if they are different from those we wanted to measure. To simplify our example, and focus on confounding and selection bias, we omit a discussion of measurement nodes in this paper and refer the reader to (Hernán and Robins 2020) for a detailed discussion on measurement bias.

We have now specified our assumptions about the data generation process. Since we are using simulated data in this paper, we will build these assumptions into our simulation, which we discuss in the next section. In a real-world example, these assumptions are used to select which variables to adjust for to remove bias in the estimate of the average causal effect of childcare attendance on conduct problems in school (e.g., (Rey-Guerra et al. 2022)). We demonstrate this process in Section 5.

# 4 Data

The analysis in this paper is based on simulated data. The advantage of using a simulated dataset is that we know what the true causal effect should be, so we know that any deviations from that indicate bias. We will generate data which is consistent with the causal diagram in Figure 7. We will, in fact, generate two datasets. Recall that in the description of the case study in Section 3, we said that our data came from children who attended a certain weekend play group. We will first generate the data including children that both do and do not attend the weekend playgroup. The second dataset will then restrict to only the children who attend the weekend playgroup. Comparing the results of the two datasets will demonstrate the selection bias in the second dataset.

The data we are generating will be *no effect data*. In other words, we will delete the edge from the treatment "childcare" to the outcome "conduct problems (school)," so that there is no causal effect between treatment and outcome. This means that when estimating the average causal effect in Section 6, we know that any deviation from no causal effect indicates a bias in our analysis.

The causal diagram in Figure 7 specifies the conditional dependencies of the random variables represented by nodes in the graph. To simulate the data, we choose the type of random variable and the exact relationships between the parameters. Since all the variables are binary, we take them to be Bernoulli random variables $Bern(p)$, which is true with probability $p$. Our task in simulating the data is to specify the probabilities.

We begin with the variables genetic effect $G$ and parental education $E$, since they have no incoming arrows, indicating their probability is constant

$$G{\sim}Bern(0.1)$$

$$E{\sim}Bern(0.9)$$

The constant probabilities, and those following, were chosen to inflate the biases we will demonstrate when analysing the data in Section 6. We can then start "building" the other random variables, based on these. For example, the causal diagram Figure 7 indicates that positive carer-child interactions $I$ depend only on parent education $E$, so we can simulate it as

$$I{\sim}Bern(0.1 + 0.85E)$$

Note that the Bernoulli random variable $E$ takes values zero or one, so the probability define above can take values 0.1 or 0.9. This means that in our simulated data, children who have highly educated parents have a higher probability of positive child-carer interactions.

The rest of the random variable definitions can be formed using the same process. We summarise the definitions in Table 1. We drew data from these distributions using the built-in function rbinom in the programming language R, which generates a Bernoulli random variable when the number of trials is set to one. We simulated a dataset of one million observations and this comprises of our first dataset, including children that both attended and did not attend the weekend playgroup. To make the second dataset, we only include the data for children who did attend the weekend playgroup, leaving us with 696,539 observations. The code is available in the supplementary material of this paper.

**Table 1 Definitions of the random variables used to simulate data for the simulated case study on the impact of childcare attendance on a child's behavioural development, measured by whether they have conduct problems at school**

| VARIABLE NAME | TYPE | SYMBOL | DISTRIBUTION |
|---|---|---|---|
| Genetic effect | Confounder | $G$ | $G{\sim}Bern(0.1)$ |
| Parent education | Confounder | $E$ | $E{\sim}Bern(0.9)$ |
| Positive carer-child interactions | Confounder | $I$ | $I\|E{\sim}Bern(0.1 + 0.85E)$ |
| Conduct problems (entry) | Confounder | $C_e$ | $C_e\|I,G \sim Bern(0.65 - 0.3E - 0.3I + 0.3G)$ |
| Attended childcare | Treatment | $A$ | $A\|C_e \sim Bern(0.25 + 0.5C_e)$ |
| Weekend playgroup | Collider | $P$ | $P\|A,E \sim Bern(0.1 + 0.34A + 0.54E)$ |
| Conduct problems (school) | Outcome | $C_s$ | $C_s\|E,C_e \sim Bern(0.65 - 0.3E + 0.3C_e - 0.3I)$ |

Having generated a synthetic dataset, we can proceed to analyse this data. Recall that our simulated data is *no effect data*, meaning that it has been constructed so that there is no causal effect of the treatment on the outcome. This means that any observed causal in the estimate of the average causal effect will indicate bias in our estimate. However, before demonstrating these biases we need to introduce some adjustment methods, which we will do in the next section.

# 5 Methods

The main idea behind the statistical methods used in causal analysis is to make the observational data we have access to behave, as close as we can, to data from which we can draw causal inference. Ideally, we would have the outcome of each person in our study both under treatment and control conditions. This (fictional) data would allow us to compute the individual causal effect for each person in the study. However, since it is impossible to collect such data (as it would involve treating the person, measuring the outcome, going back in time, not treating them and then measuring the second outcome), the next best option is a randomised control trial (RCT). In an RCT, people are randomised to treatment or not, so we can calculate the average causal effect by comparing the outcomes from the two populations (treated and control). When the study uses observational data, we use statistical methods to approximate the average causal effect we would have obtained from doing a randomised control trial.

In this section, we describe the three statistical adjustment methods used to obtain our results. These three methods are called outcome regression, G-computation and inverse probability weighting (IPW). We have chosen these methods as they are the most basic described by (Hernán and Robins 2020). Outcome regression is seen as a starting point for adjusting for confounding, whereas G-computation and IPW are more effecting for adjusting for both confounding and selection bias. All these methods are based on generalised linear models.

The statistical models adjust for the variables as determined by the causal diagram, so we describe which variables to adjust for in the case study in the second part of this section.

## 5.1 Statistical adjustment methods

Outcome regression makes a model for the outcome in terms of the treatment and any variables selected from the causal diagram to adjust for confounding. The idea behind outcome regression is that we are stratifying the data by the variables needed to adjust for confounding. By comparing the treated and control groups within levels of these variables, the populations are exchangeable, so we can calculate the average treatment effect. For a binary outcome, logarithmic binomial regression is used on the outcome

$$\log \mathrm{P}(Y = 1) = \theta_0 + \theta_1 A + \theta_2 L + \varepsilon$$

where P indicates the probability of, $Y$ is the outcome, $A$ the treatment, $L$ represents one adjustment variable, but any number can be added, and ε represents the error. The exponential of the coefficient of the treatment $exp(\theta_1)$ is the causal risk ratio.

In G-computation, the causal analysis is treated as the missing data problem. Returning to the idea that ideally, we could measure the outcome for each participant with and without treatment, G-computation creates a model for the outcome to estimate the outcome for each participant for treatment they did not receive, i.e. for every participant that received the treatment, it estimates their outcome where they not treated and vice versa for control participants. This gives an outcome under both treatment and no treatment for all participants, so the average causal effect can be calculated by taking their difference or ratio. In the case of a binary outcome, the causal risk ratio can be calculated by averaging the outcome in the treated and control groups and dividing them. For a binary outcome, logistic regression can be used, since we are not interpreting the coefficient of the treatment: we are using the model to predict the outcome to fill in missing data.

$$\log \frac{Y}{1 - Y} = \theta_0 + \theta_1 A + \theta_2 L + \varepsilon$$

Interaction terms between the treatment and adjustment variables can be added so that the treatment effect is not assumed to be the same within all levels of covariates. This cannot be done in outcome regression.

Inverse probability weighting (IPW) creates a pseudo-population by using a model for the treatment to reweight the data to balance it with respect to the adjustment variables chosen from the causal diagram. In this way, the data mimics a population where the biases identified in the causal diagram have been removed. There are two main steps. The first is to calculate the weights by creating a model for the binary treatment using logistic regression

$$\log \frac{A}{1 - A} = \theta_0 + \theta_1 L + \varepsilon$$

This model gives the probability of being treated based on the adjustment variables. The inverse probability weight is then calculated for each participant as the inverse of the probability of treatment if they were treated, and the inverse of the probability they were not treated if they were not. Having calculated the weights, the second step is to calculate the average causal effect, which in the case of a binary treatment is the causal risk ratio given by $exp(\theta_1)$ using logarithmic binomial regression

$$\log Y = \theta_0 + \theta_1 A + \varepsilon$$

solved by a weighted least squares method, where the weights are the inverse probabilities calculated in the first step.

An advantage of G-computation and IPW over outcome regression is that they can be used to just for implicit selection bias that a dataset can contain when it selects on a certain part of the population. We will demonstrate this in our simulated case study by showing that both G-computation and IPW are effective at removing the selection bias introduced by collecting the dataset from a group of children who attend a certain weekend playgroup.

Since G-computation and IPW have different assumptions, the first uses a model of the outcome and the second a model of the treatment, applying both methods during an analysis and comparing the results is a useful check. If both models are well-specified, they should give similar results, so if their results are wildly different it is an indication that one or both models are badly specified.

In this section we gave a brief overview of the three adjustment methods we used to obtain the results in the next section. We will compare their respective estimates in the discussion in Section 7 after presenting the results.

## 5.2 Variable selection

To estimate the causal effect of attending childcare on cognitive behaviour in school, we need to adjust for all variables that remove confounding and selection bias. In Figure 7, there is a back-door path through which noncausal association flows from treatment to outcome through the variable conduct problems (entry). Note that all the confounding can be adjusted for just through this variable, so in this (simple) case study, adjusting for conduct problems removes all the confounding bias.

The path between treatment and outcome that goes through the collider weekend play group is blocked, so it does not cause any bias. However, we demonstrate this structure for two reasons. Firstly, a naïve statistical analysis which included all measured variables in the data, such as weekend play group, would introduce selection bias into the estimate, because adjusting for a collider opens the path of association. The second reason is that if our data set were a convenience sample, taken from data collected by a weekend play group, this would also introduce selection bias into the estimate of the causal effect. With data generated as described by Figure 7, this bias could be removed by adjusting for parent education as well. These assertions are demonstrated quantitatively in the results.

# 6 Results

Here we present tables of results of various analysis on the simulated data described in Section 4. The analysis was done in R (R Core Team 2022) using built in regression functions and code for G-computation (Shepherd [2022] 2022). The data was generated so that the treatment has no causal effect on the outcome, meaning that a risk ratio away from one indicates a bias in the estimate. We undertake multiple analyses of the same data to demonstrate two points

1. How the inclusion or exclusion of variables from the adjustment procedure can remove or introduce bias. In general, adjusting for confounders removes bias and adjusting for colliders introduces bias.
2. To compare outcome regression, G-computation and inverse probability weighting (IPW), and to demonstrate that outcome regression is inferior as it cannot be used to remove implicit selection bias.

From the theory of causal diagrams, we can determine which variables should be adjusted for (to remove bias) and which variables should not be adjusted for (as they would introduce bias). In the causal diagram representing the observational data for the case study of the impact of childcare attendance on conduct problems in shown in Figure 7, we can see that there is confounding bias through the node “conduct problems (entry)” and that adjusting for this variable removes all confounding bias in our study. In short, we expect a risk ratio of one when adjusting for “conduct problems (entry)”, as the true result is that there is no effect of the treatment on the outcome. However, adjusting for “weekend playgroup” would open a back-door path between treatment and outcome and introduce bias, since “weekend playgroup” is a collider. This type of bias is referred to as selection bias, as it can also be introduced when data is restricted to a certain subset of the population, where that subset takes one particular value of a collider. In this simulated case study, that would occur if the data set only contained children who attended the weekend play group. This means that the collider “weekend play group” would be adjusted for implicitly and would introduce bias into the causal effect estimate. This bias can be mitigated, however, by adjusting for parent education, as it would close that back-door path between treatment and outcome.

This behaviour is encapsulated in four scenarios, where we estimate the causal effect of childcare attendance on conduct problems (school). The first two use the full data set

1. No adjustment and adjusting for the confounder conduct problems (entry),
2. Adjusting for the collider weekend playgroup,

The second two are on a reduced data set, only including children who attended the weekend play group,

3. Adjusting only for confounder conduct problems (entry),
4. Adjusting for both conduct problems (entry) and parent education.

The results are presented in Tables 2-5. The analysis was done in R using the glm function, and the confidence interval for the G-computation is computed using a bootstrap method.

The first analysis, shown in Table 2, demonstrates how to deal with confounding bias. The first row of the table shows that that without adjustment, a biased estimate of the causal effect will be obtained. The estimate was obtained using a general linear model with only the treatment included in the model, but the same point estimate would have been obtained by dividing the proportion of children with conduct issues out of those that did attend childcare by the proportion of children with conduct issues out of those that did not:

$$\frac{(\#\text{children with conduct problems who attended childcare})/(\#\text{children in childcare})}{(\#\text{children with conduct problems who did not attend childcare})/(\#\text{children not in childcare})}$$

This direct comparison of the conduct problems in school of children who did or did not attend childcare gives a biased risk ratio greater than one, falsely indicating that children who attend childcare have more conduct problems than those who do not. Looking at the causal diagram, we know that this is confounding bias through a back-door path through the node “conduct problems (entry)”. The last three rows of Table 2 confirm quantitatively that adjusting for conduct problems (entry) removes the bias and gives the known, true risk ratio of one. All three adjustment methods perform well in this example, where the only bias is confounding bias.

**Table 2 Estimation of the average causal effect of childcare attendance on conduct problems in school, on simulated data where there is no causal effect (giving a true risk ratio of one). With no adjustment, the result is biased, but all three adjustment methods give a good estimate when adjusting for confounder conduct problems on entry.**

| MODEL | ADJUSTMENT VARIABLE(S) | RISK RATIO | CONFIDENCE INTERVAL |
|---|---|---|---|
| No adjustment | - | 2.4129 | (2.3897, 2.4363) |
| Outcome regression | Conduct problems (entry) | 1.0006 | (0.9896, 1.0118) |
| G-computation | Conduct problems (entry) | 1.0006 | (0.9924, 1.0086) |
| IPW | Conduct problems (entry) | 1.0006 | (0.9938, 1.0075) |

The second analysis, shown in Table 3, demonstrates how including a collider in the adjustment procedure can introduce bias into the estimate of the causal effect, known as selection bias. For this reason, it is important to choose variables carefully when conducting a causal analysis, only choosing those that are expected to be confounders, rather than simply including all available variables in the analysis.

**Table 3 Estimation of the average causal effect of childcare attendance on conduct problems in school, on simulated data where there is no causal effect (giving a true risk ratio of one). As well as adjusting for confounder conduct problems (entry), here we adjust for the collider weekend play group, which introduces bias with all three methods. This is because adjusting for the collider opens a non-causal path between treatment and outcome.**

| MODEL | ADJUSTMENT VARIABLE(S) | RISK RATIO | CONFIDENCE INTERVAL |
|---|---|---|---|
| Outcome regression | Conduct problems (entry), Weekend play group | 1.2453 | (1.2306, 1.26018) |
| G-computation | Conduct problems (entry), Weekend play group | 1.2905 | (1.2758, 1.3029) |
| IPW | Conduct problems (entry), Weekend play group | 1.4097 | (1.4006, 1.4188) |

The final two analyses, in Tables 4 and 5, demonstrate the case where a collider is adjusted for implicitly by the data being a selected part of the population, which happens to correspond to a collider. The data is selected to contain only entries from children that attend a certain weekend play group. From the causal diagram, we can see that adjusting for the node “weekend play group” should introduce bias, and Table 4 shows that it does, since all the methods give a risk ratio away from one. It is not always possible to remedy this type of selection bias, but for data generated from our causal diagram it can, by blocking the flow of non-causal association by adjusting for parent education. Table 5 demonstrates how both G-computation and IPW give a good estimate of the expected risk ratio of one when including parent education in the model. However, outcome regression does not perform as well, which is one of the reasons it is not a recommended method.

**Table 4 Estimation of the average causal effect of childcare attendance on conduct problems in school, on simulated data where there is no causal effect (giving a true risk ratio of one), but where the data has been restricted to only children that attend a weekend play group. The risk ratios from all methods are biased due to selection bias, since weekend play group is a collider and all children in the data set attend that play group.**

| MODEL | ADJUSTMENT VARIABLE(S) | RISK RATIO | CONFIDENCE INTERVAL |
|---|---|---|---|
| Outcome regression | Conduct problems (entry) | 1.1409 | (1.1230, 1.1590) |
| G-computation | Conduct problems (entry) | 1.1273 | (1.1116, 1.1429) |
| IPW | Conduct problems (entry) | 1.1485 | (1.1380, 1.1592) |

**Table 5 Estimation of the average causal effect of childcare attendance on conduct problems in school, on simulated data where there is no causal effect (giving a true risk ratio of one), but where the data has been restricted to only children that attend a weekend play group. In each of the models, the selection bias has been mitigated by adjusting for parent education. Note that outcome regression performs worse than the other two methods in reducing this bias, which is one of the reasons G-computation and IPW are preferred methods.**

| MODEL | ADJUSTMENT VARIABLE(S) | RISK RATIO | CONFIDENCE INTERVAL |
|---|---|---|---|
| Outcome regression | Conduct problems (entry), Parent education | 1.0426 | (1.0259, 1.0597) |
| G-computation | Conduct problems (entry), Parent education | 1.0119 | (0.9954, 1.0107) |
| IPW | Conduct problems (entry), Parent education | 1.0092 | (1.0001, 1.0185) |

# 7 Discussion

A resounding issue for researchers, governments, and educational stakeholders is drawing causal inferences from observational data to inform understanding, policy, and practice. In this article we highlighted the latest developments in using causal diagrams and statistical methodology to best infer causality and communicate assumptions underlying this inference.

Our simulated case study provided a quantitative demonstration of confounding and selection bias. We illustrated how a causal diagram and statistical adjustment can be used to both remove bias when the correct variables are adjusted for, and inadvertently increase bias if the wrong variables are included. Although many researchers are familiar with adjusting for confounding variables, this simulation showed that including too many variables can increase the likelihood of identifying a false causal effect (Robins and Greenland 1986). Likewise, although the inclusion of covariates is routinely justified by an underlying understanding of the system being studied, explicitly drawing the causal diagram communicates the assumptions of the analysis. Given the importance of evaluating model assumptions and specification, the recommendation from this simulation is to provide casual diagrams when undertaking observational analyses (Grosz, Rohrer, and Thoemmes 2020).

A secondary aim of the simulation was to compare three adjustment methods: outcome regression, G-computation and IPW. Both G-computation and IPW perform well in the simulated case study, though are limited by adjusting for the wrong variable (collider bias) or in the presence of unadjusted selection bias. Outcome regression performed well when adjusting for confounding but was also insufficient to overcome collider bias and comparatively worse in the presence of selection bias even when using a secondary indicator. The implication from these results are that analysts should also consider broadening their expertise to include G-computation and IPW (amongst other weighting methodology; (Guo and Fraser 2014; Hernán and Robins 2020; Bang and Robins 2005)) alongside typical regression approaches. Convergence across multiple methods will improve confidence in research findings, though presenting the assumptions underlying the analysis remains paramount for transparency and accumulating scientific knowledge.

Although we focused on casual associations between variables, there are also design elements of longitudinal analysis that can improve causal inferences that we did not explore. For example, within-person designs can be used to adjust for time invariant unobserved heterogeneity provided sufficient variation in both the outcome and variable exist at the person level across time (e.g., fixed effects and hybrid models; (Hamaker and Muthén 2020)). However, these longitudinal designs still require assumptions regarding the data generating process in terms of selection bias and time-varying confounding, thus causal diagrams and methods are equally as applicable in these instances (Hernán and Robins 2020).

Finally, while we have made a strong case for the use of causal inference methods on administrative data, there are some limitations. First, the latest causal inference techniques are not yet widespread enough to be taught in mainstream undergraduate statistics courses, so analysts need to learn the theory behind these methods before applying them to their studies. Second, a crucial part of applying these techniques is drawing the causal diagram which requires understanding the system of interest. However, even in established fields there may be limited or contested consensus of how the system functions at the individual variable level and finding relevant subject matter experts may be a challenge for data analysts working to support policy development. Causal discovery, where the causal diagram is learned from the data (Vowels, Camgoz, and Bowden 2022; Nogueira et al. 2022) could be useful in these cases, though ideally in combination with subject matter expertise. Third, the variables considered important by researchers for system and question of interest may not necessarily be captured by administrative or observational data sets. It can be important, therefore, to consider using causal diagrams to assist in the design of data collection to augment administrative data.

# 8 Conclusion

In this paper, we demonstrated how improvements in causal inference techniques can be used to inform decisions in policy making using administrative and observational data. We did this by introducing causal inference concepts and using a simulated case study to demonstrate both how to do an analysis and common types of bias in observational studies. We showed that confounders introduce bias which can be removed by adjusting for them. We also demonstrated that selection bias can occur in two ways. The first is by a collider variable being included in the model. This indicates a key difference between causal and predictive modelling: the variable selection process. In predictive modelling the aim is to include variables that predict the outcome well. In causal modelling, it is important to only include the confounders required for adjustment. In general, including extra variables might have no effect on the estimate of the causal effect, but if such a variable is a collider it can introduce bias. The second way that selection bias can occur is by selecting a certain part of the population which implicitly adjusts for a collider, as with restricting to only children that attended a certain weekend play group in our simulated case study. While this kind of selection bias cannot always be removed, we showed how both G-computation and IPW do well at removing this bias in our case study, and that outcome regression did not perform as well.

Causal inference techniques have great potential to maximise the value of administrative data sets by improving estimates of the impact of policy changes on the community. This process needs to include collaboration between subject matter experts and data analysts, to ensure that state-of-the-art statistical methods are backed by relevant understanding of the system to create assumptions of how the data was generated. Using these causal methods will lead to better and more transparent research and decisions that can improve public life.

# References

Almgerbi, Mohamad, Andrea De Mauro, Adham Kahlawi, and Valentina Poggioni. 2022. 'A Systematic Review of Data Analytics Job Requirements and Online-Courses'. *Journal of Computer Information Systems* 62 (2): 422–34. https://doi.org/10.1080/08874417.2021.1971579.

Bang, Heejung, and James M. Robins. 2005. 'Doubly Robust Estimation in Missing Data and Causal Inference Models'. *Biometrics* 61 (4): 962–73. https://doi.org/10.1111/j.1541-0420.2005.00377.x.

Connell, Arin M., and Sherryl H. Goodman. 2002. 'The Association between Psychopathology in Fathers versus Mothers and Children's Internalizing and Externalizing Behavior Problems: A Meta-Analysis'. *Psychological Bulletin* 128: 746–73. https://doi.org/10.1037/0033-2909.128.5.746.

Davis, Elspeth, Areana Eivers, and Karen Thorpe. 2012. 'Is Quality More Important If You're Quirky? A Review of the Literature on Differential Susceptibility to Childcare Environments'. *Australasian Journal of Early Childhood* 37 (4): 99–106. https://doi.org/10.1177/183693911203700414.

Egert, Franziska, Ruben G. Fukkink, and Andrea G. Eckhardt. 2018. 'Impact of In-Service Professional Development Programs for Early Childhood Teachers on Quality Ratings and Child Outcomes: A Meta-Analysis'. *Review of Educational Research* 88 (3): 401–33. https://doi.org/10.3102/0034654317751918.

Grosz, Michael P., Julia M. Rohrer, and Felix Thoemmes. 2020. 'The Taboo Against Explicit Causal Inference in Nonexperimental Psychology'. *Perspectives on Psychological Science* 15 (5): 1243–55. https://doi.org/10.1177/1745691620921521.

Guo, Shenyang, and Mark W. Fraser. 2014. *Propensity Score Analysis: Statistical Methods and Applications*. Vol. 11. SAGE publications.

Hamaker, Ellen L., and Bengt Muthén. 2020. 'The Fixed versus Random Effects Debate and How It Relates to Centering in Multilevel Modeling'. *Psychological Methods* 25: 365–79. https://doi.org/10.1037/met0000239.

Henry, Regina, and Santosh Venkatraman. 2015. 'Big Data Analytics the Next Big Learning Opportunity'. *Journal of Management Information and Decision Sciences* 18 (2): 17–29.

Hernán, Miguel A., dir. 2019. *Draw Your Assumptions Before Your Conclusions*. edX online course.

Hernán, Miguel A., John Hsu, and Brian Healy. 2019. 'A Second Chance to Get Causal Inference Right: A Classification of Data Science Tasks'. *CHANCE* 32 (1): 42–49. https://doi.org/10.1080/09332480.2019.1579578.

Hernán, Miguel A., and James M. Robins. 2020. *Causal Inference: What If.* Boca Raton: Chapman & Hall/CRC.

Letourneau, Nicole Lyn, Linda Duffett-Leger, Leah Levac, Barry Watson, and Catherine Young-Morris. 2013. 'Socioeconomic Status and Child Development: A Meta-Analysis'. *Journal of Emotional and Behavioral Disorders* 21 (3): 211–24. https://doi.org/10.1177/1063426611421007.

Lovejoy, M. Christine, Robert Weis, Elizabeth O'Hare, and Elizabeth C. Rubin. 1999. 'Development and Initial Validation of the Parent Behavior Inventory.' *Psychological Assessment* 11 (4): 534.

May, Felix, Tamsin Ford, Astrid Janssens, Tamsin Newlove-Delgado, Abigail Emma Russell, Javid Salim, Obioha C. Ukoumunne, and Rachel Hayes. 2021. 'Attainment, Attendance, and School Difficulties in UK Primary Schoolchildren with Probable ADHD'. *British Journal of Educational Psychology* 91 (1): 442–62. https://doi.org/10.1111/bjep.12375.

Moon, Deborah J., Jeri L. Damman, and Aly Romero. 2020. 'The Effects of Primary Care–Based Parenting Interventions on Parenting and Child Behavioral Outcomes: A Systematic Review'. *Trauma, Violence, & Abuse* 21 (4): 706–24. https://doi.org/10.1177/1524838018774424.

Nogueira, Ana Rita, Andrea Pugnana, Salvatore Ruggieri, Dino Pedreschi, and João Gama. 2022. 'Methods and Tools for Causal Discovery and Causal Inference'. *WIREs Data Mining and Knowledge Discovery* 12 (2): e1449. https://doi.org/10.1002/widm.1449.

Pearl, Judea. 1995. 'Causal Diagrams for Empirical Research'. *Biometrika* 82 (4): 669–88. https://doi.org/10.1093/biomet/82.4.669.

———. 2009. *Causality*. Cambridge University Press.

Pinquart, Martin. 2017. 'Associations of Parenting Dimensions and Styles with Externalizing Problems of Children and Adolescents: An Updated Meta-Analysis'. *Developmental Psychology* 53: 873–932. https://doi.org/10.1037/dev0000295.

R Core Team. 2022. *R: A Language and Environment for Statistical Computing*. Vienna, Austria: R Foundation for Statistical Computing. https://www.R-project.org/.

Rankin, Peter Sheldon, Sally Staton, Azhar Hussain Potia, Sandy Houen, and Karen Thorpe. 2022. 'Emotional Quality of Early Education Programs Improves Language Learning: A within-Child across Context Design'. *Child Development* 93 (6): 1680–97. https://doi.org/10.1111/cdev.13811.

Rey-Guerra, Catalina, Henrik D. Zachrisson, Eric Dearing, Daniel Berry, Susanne Kuger, Margaret R. Burchinal, Ane Nærde, Thomas van Huizen, and Sylvana M. Côté. 2022. 'Do More Hours in Center-Based Care Cause More Externalizing Problems? A Cross-National Replication Study'. *Child Development* n/a (n/a). https://doi.org/10.1111/cdev.13871.

Robins, James M., and Sander Greenland. 1986. 'The Role of Model Selection in Causal Inference from Nonexperimental Data'. *American Journal of Epidemiology* 123 (3): 392–402. https://doi.org/10.1093/oxfordjournals.aje.a114254.

Rutter, Michael. 2006. *Genes and Behavior: Nature-Nurture Interplay Explained*. Genes and Behavior: Nature-Nurture Interplay Explained. Malden: Blackwell Publishing.

Shepherd, Daisy. (2022) 2022. 'G-Computation'. R. https://github.com/daisyshep/G-computation.

Textor, Johannes, Benito van der Zander, Mark S. Gilthorpe, Maciej Liśkiewicz, and George TH Ellison. 2016. 'Robust Causal Inference Using Directed Acyclic Graphs: The R Package "Dagitty"'. *International Journal of Epidemiology* 45 (6): 1887–94. https://doi.org/10.1093/ije/dyw341.

Vowels, Matthew J., Necati Cihan Camgoz, and Richard Bowden. 2022. 'D'ya Like DAGs? A Survey on Structure Learning and Causal Discovery'. *ACM Computing Surveys* 55 (4): 82:1-82:36. https://doi.org/10.1145/3527154.

Weeland, Joyce, Geertjan Overbeek, Bram Orobio de Castro, and Walter Matthys. 2015. 'Underlying Mechanisms of Gene–Environment Interactions in Externalizing Behavior: A Systematic Review and Search for Theoretical Mechanisms'. *Clinical Child and Family Psychology Review* 18 (4): 413–42. https://doi.org/10.1007/s10567-015-0196-4.

## A.1 Simulation of basic causal diagrams

In this appendix we simulate data based on the three basic causal diagrams and demonstrate how adjusting for common causes and mediators blocks association and adjusting for colliders induces association.

### A.1.1 Common cause

A common cause between two nodes in a causal diagram induces noncausal association between those nodes. Figure 8 gives an example where node C is a common cause of nodes A and B. In data generated as described by this diagram, the confounder C induces noncausal association between variables A and B. We will demonstrate this by simulating data based on the diagram in Figure 8, where all the variables are binary.

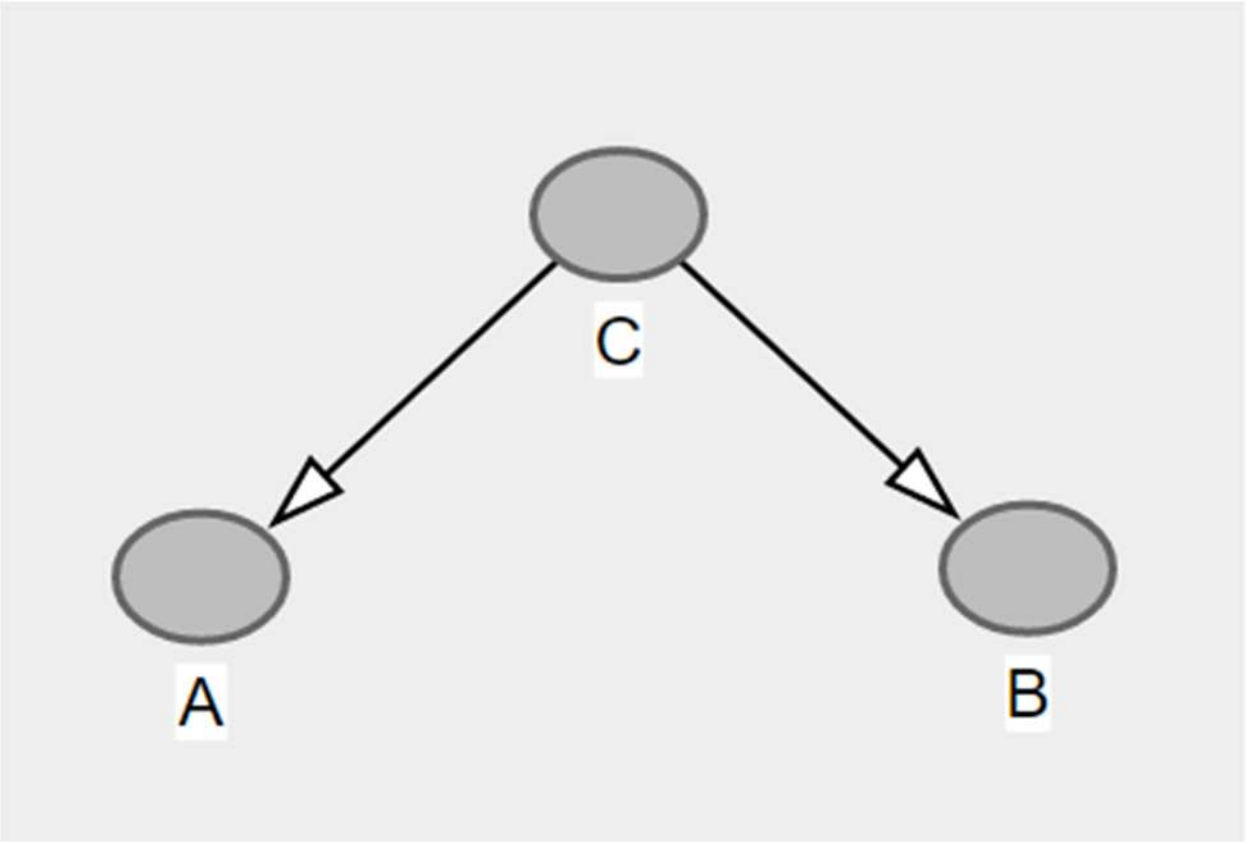

Figure 8 Node C is a confounder or common cause of nodes A and B

Since node C has no arrows pointing into it, it depends on no other variables in the graph, so we take it to have a binomial distribution with constant probability of success.

$$C \sim \text{Binom}(0.5)$$

The distributions of A and B, however, depend on the distribution of C

$$A \sim \text{Binom}(0.25 + 0.5C)$$
$$B \sim \text{Binom}(0.25 + 0.5C).$$

We have chosen the probabilities such that they fall between zero and one for both values of $C \in \{0,1\}$. We simulated data in R based on these distributions, drawing ten thousand data points.

Our aim is to study the association between variables A and B in this dataset. We did this by fitting a Poisson generalised linear model with a log link function, taking B as the response and A as a covariate. If there were no association between the variables, we would expect the exponential of the coefficient of A to be one. We refer to this value as the *risk ratio*. A risk ratio away from one indicates an association between variables A and B.

As expected, due to the common cause C, the model which does not adjust for C gives a biased risk ratio, i.e. the value of the risk ratio is away from one. However, adjusting for common cause C gives the expected result of a risk ratio of one.

Table 6 Estimation of risk ratio between A and B in Figure 8. The random variables A and B are independent, so we expect a risk ratio of 1. The correct result is obtained by adjusting for common cause C, shown in the second row of the table. Failing to adjust for C gives a biased result shown in the first row.

| MODEL | RISK RATIO | 95% CONFIDENCE INTERVAL | PATH BETWEEN A AND B |
|---|---|---|---|
| $\boldsymbol{A \sim B}$ | 1.696 | (1.602, 1.795) | Open |
| $\boldsymbol{A \sim B + C}$ | 1.031 | (0.967, 1.099) | Closed |

### A.1.2 Mediator

A mediator between two nodes in a causal diagram allows association to flow through it. However, adjusting for a mediator blocks the flow of association. Therefore, in Figure 9, we expect (causal) association between nodes A and B, because association flows from A to B through C in the direction of the arrows. Adjusting for C would block this flow of association and would give a risk ratio of one, introducing bias and indicating no association between A and B. We will demonstrate this by simulating data based on Figure 9.

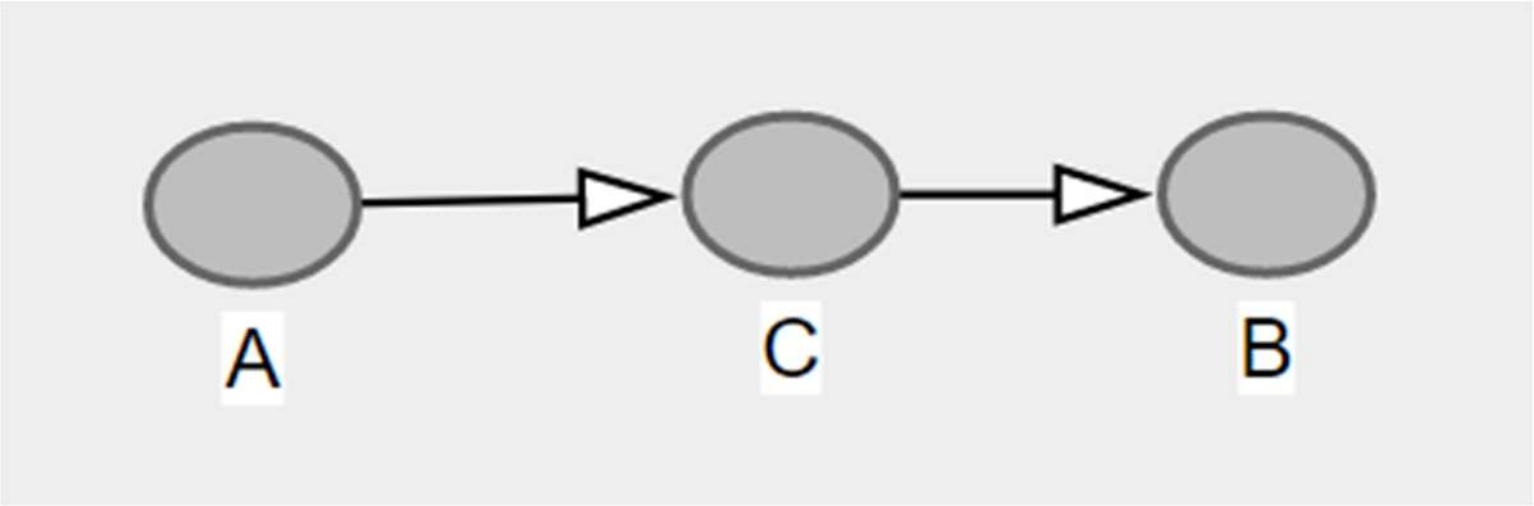

**Figure 9 Node C is a mediator between nodes A and B**

The following random variables could be represented by the variables in Figure 9

$$A \sim \text{Binom}(0.5)$$

$$C \sim \text{Binom}(0.25 + 0.5A)$$

$$B \sim \text{Binom}(0.25 + 0.5C)$$

We simulated ten thousand data points in R based on these distributions and use them to model the data and obtain an estimate of the risk ratio, as shown in Table 7.

**Table 7 Estimation of the risk ratio between nodes A and B in Figure 9. The first row demonstrates the (causal) association between A and B since the risk ratio is different from one. The second row shows that by adjusting for mediator C, the association between A and B is blocked, giving a risk ratio near one.**

| MODEL | RISK RATIO | 95% CONFIDENCE INTERVAL | PATH BETWEEN A AND B |
|---|---|---|---|
| $\boldsymbol{A \sim B}$ | 1.656 | (1.565, 1.754) | Open |
| $\boldsymbol{A \sim B + C}$ | 0.983 | (0.922, 1.047) | Closed |

As expected, without adjusting for mediator C, there is an association between A and B, as evidenced by a risk ratio different from one. However, by adjusting for the mediator C, the association is blocked, and the risk ratio is near one.

### A.1.3 Collider

A collider is a common effect of two nodes. When it is not adjusted for, a collider blocks the (noncausal) flow of association between two nodes. However, adjusting for a collider unblocks (opens) the path and introduces bias. In Figure 10, node C is a collider of nodes A and B. We will generate data that could be represented by this diagram to demonstrate how adjusting for a collider introduces bias by opening the path of association between nodes A and B.

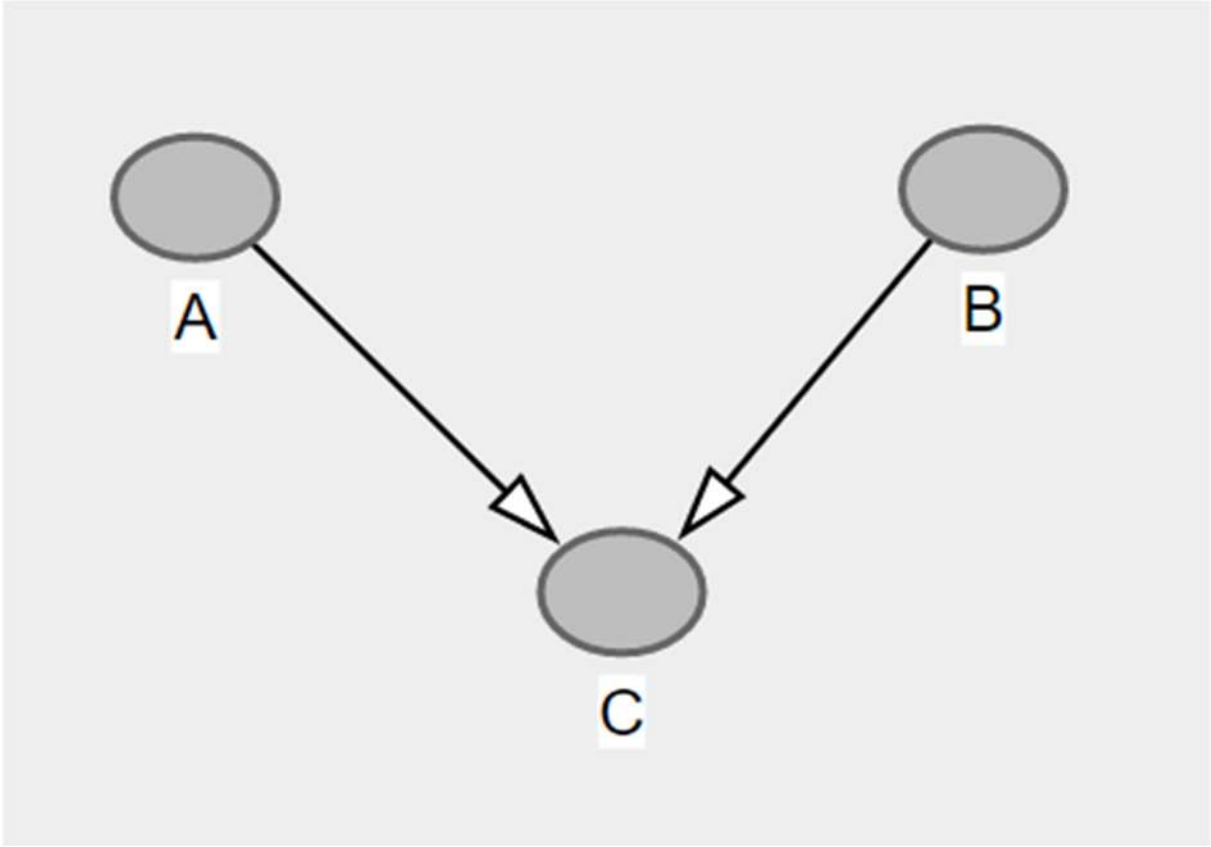

Figure 10 Node C is a collider or common effect of nodes A and B

Some random variables that could be represented by Figure 10 are

$$A \sim \text{Binom}(0.1)$$

$$B \sim \text{Binom}(0.1)$$

$$C \sim \text{Binom}(0.15 \; + \; 0.4A + 0.4B)$$

We drew ten thousand data points from these distributions and then estimated the risk ratio, see Table 8. The risk ratio in the first row, obtained by not adjusting for collider C, gives an estimate close to one, since random variables A and B are independent. However, the second row shows that by adjusting for the collider C, bias is introduced, and the risk ratio indicates an association since its value is not close to one (and the confidence interval does not include one).

**Table 8 Estimation of the risk ratio between nodes A and B in Figure 10. The first row represents the correct lack of association between nodes A and B, demonstrated by the risk ratio near one. The second row demonstrates that bias is introduced by adjusting for collider C, since the risk ratio is not close to one.**

| MODEL | RISK RATIO | 95% CONFIDENCE INTERVAL | PATH BETWEEN A AND B |
|---|---|---|---|
| $\boldsymbol{A \sim B}$ | 1.093 | (0.883, 1.336) | Closed |
| $\boldsymbol{A \sim B + C}$ | 0.546 | (0.439, 0.670) | Open |

**As Australia's national science agency and innovation catalyst, CSIRO is solving the greatest challenges through innovative science and technology.**

CSIRO. Unlocking a better future for everyone.

**Contact us**

1300 363 400
+61 3 9545 2176
csiroenquiries@csiro.au
csiro.au

**For further information**

D**ata61**
Elena Tartaglia
+61 3 9545 8548
elena.tartaglia@data61.csiro.au
csiro.au/data61